# Fast drying of high-alumina MgO-bonded refractory castables

V. S. Pinto[(1)], D. S. Fini[(1)], V. C. Miguel[(1)], V. C. Pandolfelli[(1)], M. H. Moreira[(1)], T. Venâncio[(2)], A. P. Luz[(1)*]

[(1)] Federal University of São Carlos, Materials Engineering Department (DEMa)
[(2)] Federal University of São Carlos, Chemistry Department (DEQ),
Rod. Washington Luiz, km 235, São Carlos, SP, 13565-905, Brazil.

*Corresponding author at: t*el.:* +55-16-33518601
E-mail: anapaula.light@gmail.com or analuz@ufscar.br

**Abstract**

Refractory producers face many challenges in terms of producing MgO-containing castables due to the high likelihood of magnesia to hydrate in contact with water, resulting in $Mg(OH)_2$ generation. The expansive feature of this transformation affects the performance of such refractories, as *(i)* if this hydrated phase is not accommodated in the formed microstructure, ceramic linings with cracks and low green mechanical strength will be obtained; and *(ii)* if crack-free pieces are prepared, they should present low porosity and reduced permeability, which require special attention when heating these materials. In both cases, spalling/explosion may be favored during the drying step of MgO-containing compositions. Hence, this work investigated the ability of various additives in the optimization of the drying behavior of $Al_2O_3$-MgO castables. Vibratable compositions were tested after incorporating polymeric fibers (PF), an organic salt (OAS), $SiO_2$-based additive (SM) or permeability enhancing active compound (MP) into the dry-mixtures. Various experimental measurements were performed to infer the role of the drying agents to prevent the samples' explosion and whether they would also influence other properties of the castables. As observed, OAS and MP helped to inhibit the MgO-bonded samples' explosion even under severe heating conditions (2-20°C/min) and increased their green mechanical strength and slag infiltration resistance when compared to the additive-free composition. On the other hand, the addition of polymeric fibers (PF) or silica-based compound (SM) to the formulations was not able to prevent the castables' explosion when using a high heating rate and other side effects (samples' cracking during drying at 110°C, high linear expansion and increased slag penetration during corrosion tests) could also be observed when testing these materials. Thus, the selection of suitable drying agents is a key issue, as they may allow the development of MgO-bonded castables with enhanced properties and lower spalling risk during their first thermal treatment.

**1. Introduction**

The mixing stage of refractory castables requires the addition of water or some other liquid to help mold it into the desired shape. After that, drying becomes the most important processing step, as the added water must be withdrawn from the microstructure before firing, which implicates in heat and mass transfers of $H_2O$ in liquid or vapor form [1,2]. According to Oummadi et al. [2], two main issues can be related to drying: *(i)* it requires significant energy consumption, and *(ii)* it may lead to mechanical stresses (due to pressure build-up associated with steam trapped inside the dense structure) and eventually the spalling and/or explosion of the ceramic. Thus, this subject has motivated many studies focused on better understanding the water distribution and advance of the drying front in green ceramic bodies, by using alternative experimental techniques (i.e., magnetic resonance imaging and neutron tomography) [2–5]. Moreover, other research groups have also been developing numerical models of heat and mass transfers in moist ceramics in order to predict safe and optimized heating schedules for refractory compositions [6–9].

Usually, the main castable systems that present higher likelihood to undergo spalling or explosion during their drying step are the ones containing hydraulic binders (calcium aluminate cement, hydratable alumina or magnesia), as such components react with water, giving rise to hydratable phases that decompose and release water vapor at temperatures above 100°C [10–13]. Considering that the precipitated hydrates fill in the pores of the refractory structure, resulting in materials with reduced permeability, and the vapor pressure inside the castables during heating increases exponentially as a function of the temperature (according to Antoine's equation [1]), the higher the decomposition temperature of the hydrates is, the greater the damage risk will be for such materials.

Although magnesia has been pointed out as a likely binder for alumina-based castables [12–16], there are many challenges and concerns for producing high quality and crack-free compositions in this system. In this case, the bonding effect is derived from $Mg(OH)_2$ (brucite) formation (Eq. 1) in the interstices and voids among the coarse and fine particles of the refractory. However, considering the density mismatch between MgO ($\rho_{MgO}$ = 3.5 g/cm$^3$) and brucite ($\rho_{Mg(OH)2}$ = 2.3 g/cm$^3$), this reaction can result in an expressive volumetric expansion, leading to the generation of cracks in the molded samples during curing and drying steps [12,15].

$$MgO + H_2O \rightarrow Mg(OH)_2 \qquad (1)$$

Aiming to take advantage of the MgO hydration and optimize the bonding potential of this material, some approaches are suggested in the literature, such as: *(i)* inducing faster Mg(OH)$_2$ formation before the composition setting time, when the molded castable still has enough room and some freedom to accommodate stresses [14,17], *(ii)* changing the morphology of the hydrated phase to better fit the crystal growth in the designed microstructure [18–20], and *(iii)* using hydrating agents (i.e., carboxylic acids, and others) to activate a higher number of sites and increase the Mg(OH)$_2$ nucleation rate on MgO surface, which might limit the crystal growth [12,13].

As MgO can readily react with water in its liquid or vapor form, the brucite formation can still take place during the initial stages of the drying process, resulting in an additional risk to the integrity of the refractories containing this binder. Additionally, Mg(OH)$_2$ decomposition takes place at relatively high temperatures (380-420°C) [21,22], which coupled to the reduced permeability of the resulting microstructure, may result in high steam pressure levels and, consequently, the explosion of the castable even under usual heating schedule profiles.

The drying behavior of MgO-bonded castables can be adjusted by using polymeric fibers [23], silica-based compounds [24–27], organic additives (such as lactates [28]) or favoring the formation of hydrotalcite-like phases [Mg$_x$Al$_y$(OH)$_{2x+2y}$](CO$_3$)$_{y/2}$.nH$_2$O] in the compositions [29,30]. The addition of fibers is a simple and versatile solution, but their presence decreases the flowability of the mixtures, whereas the MgO and SiO$_2$ reactions at high temperatures usually give rise to low melting point phases in alumina systems. On the other hand, the *in situ* formation of hydrotalcite-like phases seems to be a promising alternative, as such compounds commonly present their thermal decomposition in various steps in the 200-400°C range, which might decrease the spalling risk associated with the steam pressure generation [30].

Considering the aspects discussed above and the availability of novel drying agents (SioxX-Mag and Refpac MIPORE 20) designed for monolithic refractory systems, this work investigated and compared the performance of various additives in the optimization of the drying behavior of high-alumina MgO-bonded castables. Vibratable compositions were formulated and tested after the incorporation of polymeric fibers, an organic aluminum salt (aluminum salt of 2-hydroxypropanoic acid), SioxX-Mag (silica-based additive) and Refpac MIPORE 20 (permeability enhancing active compound originally designed for cement-bonded compositions), into the dry-mixtures.

## 2. Experimental

*2.1 – MgO-based suspensions*

Firstly, mixtures of dead-burnt magnesia (M-30, MgO = 98.12 wt.%, CaO = 0.80 wt.%, $SiO_2$ = 0.34 wt.% and others = 0.74 wt.%, d < 212 µm, RHI-Magnesita, Contagem, Brazil), distilled water and additives (Table 1) were prepared to investigate the likely influence of the latter compounds in the brucite formation during curing (30°C) and drying (110°C) conditions. A reference composition (MgO + distilled water) was analyzed and compared to the other prepared systems. Formic acid (AcF, molecular mass: 46 g/mol, purity = 85%, Labsynth, Diadema, Brazil) was used as the nucleation site activator, and its content (Table 1) was defined based on the MgO amount added to the mixtures, as reported by Santos Jr. [12].

Three-drying agents were also incorporated into the MgO-water mixtures: *(i)* an aluminum salt of 2-hydroxypropanoic acid (named here as organic aluminum salt = OAS, 294.18 g/mol, Quimibras Industrias Químicas S.A., Brazil) [31]; *(ii)* SioxX-Mag ($SiO_2$ = 35 wt.%, 50 wt.% $Al_2O_3$ and 15 wt.% others, Elkem, Norway, SM in Table 1), which is a silica-based drying compound designed for MgO castables [32]; and *(iii)* Refpac MIPORE 20 ($Al_2O_3$ = 39-43 wt.%, CaO = 12-15 wt.%, MgO = 16-20 wt.% and loss on ignition ~24-30%, Imerys Aluminates, France, named here as MP), which is a permeability enhancing active compound for CAC-bonded refractories [33,34]. The amount of each of these agents added to the prepared suspensions (Table 1) was based on preliminary tests carried out with alumina-magnesia based castables (see Section 2.2), keeping the same MgO/additive mass ratio.

Table 1 – Simplified compositions (wt.%) of MgO + additives prepared for the evaluation of the likely chemical reactions that could take place in the castables' matrix microstructure during curing and drying stages up to 110°C.

| Mixtures (wt.%) | MgO | Distilled water ($H_2O$) | Formic acid (AcF) | Organic aluminum salt (OAS) | SioxX-Mag (SM) | Refpac MIPORE (MP) |
|---|---|---|---|---|---|---|
| MgO-$H_2O$ | 100 | 30 | - | - | - | - |
| MgO-AcF | 100 | 30 | 5.67 | - | - | - |
| MgO-OAS* | 91.66 | 40 | 5.19 | 8.33 | - | - |
| MgO-SM | 83.33 | 30 | 4.72 | - | 16.66 | - |
| MgO-MP | 83.33 | 30 | 4.72 | - | - | 16.66 |

*Higher water demand was required while preparing the mixture containing OAS, as the ready interaction of this additive with magnesia and liquid led to the development of a viscous paste.

The dry powders (MgO + additives) were mixed with distilled water in a laboratory stirrer for approximately 5 minutes at room temperature. The aqueous suspensions were molded as cylindrical samples (30 mm x 30 mm) and kept at 30°C for 24 hours. After that, the obtained solid specimens were demolded, cut in two parts and one piece of each composition was ground, whereas the other one was dried at 110°C for another 24h and also ground after this step. The obtained powders of the hydrated materials (kept at 30°C or 110°C) were analyzed via thermogravimetric (TG) and differential scanning calorimetry (DSC) measurements in STA 449 F3 Jupiter device (Netzsch, Selb, Germany) in the 30-600°C temperature range, using 5°C/min as the heating rate, synthetic air (80% $N_2$-20% $O_2$) flow of 50 $cm^3$/min and α-$Al_2O_3$ as a correction standard. Additionally, X-ray diffraction tests were carried out using D8 Focus equipment (Bruker, Karlsruhe, Germany) and considering the following parameters: CuKα radiation [λ=1.5418 Å], nickel filter, 40 mA, 40 mV, 2θ = 4-80° and scanning step = 0.02.

MgO-OAS samples were also analyzed via $^{27}Al$ Nuclear Magnetic Resonance in solid state ($^{27}Al$-ssNMR) in order to better understand the likely interaction of OAS with MgO and water at low temperatures (30°C and 50°C for 24 hours), after drying (110°C for 24 hours) and firing (200-600°C for 2 hours) steps. The powdered samples were packed into 4 mm zirconia rotors and the measurements were carried out in a Bruker Avance III-400 device, using the magnetic field of 9.4 T (where $^{27}Al$ resonance frequency is 104.21 MHz) and spinning speed of 5 and 10 kHz. A pulse sequence consisting of one pulse of 2.5 µs (for 45-degree pulse) was employed in these experiments, using a recycle delay of 1 s between 1024 scans, an acquisition time of 138 ms for a 285 ppm spectra window.

*2.2 – Refractory castables containing different drying agents*

The second stage of this study focused on evaluating the effects of adding the selected drying agents to Al$_2$O$_3$-MgO castables. In order to obtain an optimized packing, a vibratable formulation (RefM, Table 2) was designed following the Andreasen's model [1] (Eq. 2) and considering a distribution modulus (*q*) equal to 0.26.

$$\frac{CPFT}{100} = \left(\frac{D}{D_L}\right)^q \quad (2)$$

where *CPFT* is the cumulative percent finer than *D*; *q* is the distribution modulus (which defines the coarse and fine components ratio contained in the compositions); *D* is the particle size and $D_L$ is the largest particle size of the system.

Coarse tabular alumina (d < 6 mm, Almatis, Germany), reactive and calcined aluminas (CT3000SG and CL370, Almatis, Germany) and magnesia (M30-B, d < 212 µm, RHI-Magnesita, Brazil) were the main raw materials selected for this work, as pointed out in Table 2.

Table 2 – Vibratable castable compositions based on Andreasen's packing model (q = 0.26).

| Raw materials (wt.%) | Castable compositions | | | | |
|---|---|---|---|---|---|
| | **RefM** | **RefM-PF** | **RefM-OAS** | **RefM-SM** | **RefM-MP** |
| Tabular alumina (d < 6 mm) | 83 | 83 | 83 | 83 | 83 |
| Reactive alumina (CT3000SG) | 7 | 7 | 7 | 7 | 7 |
| Calcined alumina (CL370) | 4 | 4 | 4 | 4 | 4 |
| Dead-burnt MgO (M30 < 212 µm) | 6 | 6 | 6 | 6 | 6 |
| Polymeric fibers (Emsil Dry) | - | 0.1 | - | - | - |
| Organic aluminum salt (OAS) | - | - | 0.5 | - | - |
| SioxX-Mag | - | - | - | 1.0 | - |
| Refpac MIPORE 20 | - | - | - | - | 1.0 |
| Dispersant (Castament® FS 60) | 0.2 | 0.2 | 0.2 | - | - |
| Formic acid | 0.34 | 0.34 | 0.34 | 0.34 | 0.34 |

Four drying additives, the three materials pointed out in Section 2.1 (OAS, SioxX-Mag and Refpac MIPORE 20) and a polymeric fiber (polymeric fibers, length < 6 mm, EMSIL DRY, Elkem Norway), were incorporated into the reference formulation (Table 2). The action of the fibers as a drying compound should be mainly observed after their melting and decomposition (> 300°C [35]), which might give rise to permeable paths for easier steam release. Thus, it is expected that the addition of this polymer to RefM-PF would not affect the MgO hydration reaction sequence and the samples' properties during curing and drying up to 110°C (as analyzed in Section 2.1).

The amount of each one of the evaluated additives was defined based on preliminary tests, as well as previous works presented in the literature [12,35]. As highlighted in Table 2, a total of 0.2 wt.% of a polyethylene glycol-based dispersant (Castament® FS 60, Basf, Germany) was incorporated into the dry-powders of the REFM, RefM-PF and RefM-LA formulations before their mixing step, whereas the other two compositions (RefM-SM and RefM-MP) did not require the use of such material as both drying agents (SioxX-Mag and MIPORE) already contained dispersants in their original compositions [33,36,37].

In order to better control the kinetic of the brucite nucleation and growth, a pre-activation step of the MgO particles is required as suggested by Santos Jr. [12,13]. Hence, the selected dead-burnt magnesia was mixed with distilled water and formic acid in a lab stirrer for 2 minutes. The obtained suspension was then added to the castable's dry-components during its mixing process in a rheometer device [38]. The water demand for each tested composition was adjusted in order to obtain vibratable flow values (ASTM C 1445) around 150%. After that, the prepared mixtures were molded under vibration, cured at 30°C for 24h and dried at 110°C for an additional 24h. Curing behavior of the fresh castables was also monitored via ultrasonic measurements (UltraTest device, IP-8 measuring system, Germany) to follow the propagation velocity of the ultrasonic wave in the prepared materials as a function of time and at room temperature (~22°C) for 24h.

Thermogravimetric tests were conducted in an electric furnace controlled by a proportional-integral-derivative (PID) system up to a maximum temperature of 600°C and according to different heating rates (2, 5 or 20°C/min). Cylindrical samples (50 mm x 50 mm), previously cured at 30°C for 24h, were suspended at the center of the furnace [39]. In the first set of experiments, the temperatures of the furnace and at the center of the humid specimen were monitored during each heating schedule. New tests were then carried out with similar samples, but without thermocouples to obtain the castables' mass loss under the same heating conditions. The highest heating rate (20°C/min) was chosen to analyze the castables' explosion likelihood during their first heating cycle.

Mass loss during drying was assessed through the normalized parameter $W$ (Eq. 3), which measures the cumulative water content expelled during the heating schedule per total amount of water initially contained in the humid body.

$$W(\%) = 100 \times \left( \frac{M_0 - M}{M_0 - M_f} \right) \tag{3}$$

where, $M$ is the instantaneous mass recorded at time $t_i$ during the heating stage, $M_0$ and $M_f$ are the initial and final mass of the tested specimen, respectively. The derivative of the mass loss profiles as a function of time was calculated using the Origin software (version 9, OriginLab, USA).

The cold mechanical strength of prismatic samples (150 mm x 25 mm x 25 mm) cured at 30°C for 24h and/or dried 110°C for 24h was evaluated via 3-point bending tests (ASTM C133-97), using the universal mechanical device (MTS-810, Material Test System, USA). Additionally, the apparent porosity of such materials was determined according to ASTM C380-00, using kerosene as the immersion liquid. Five samples were analyzed for each selected testing condition and the presented values consisted of the average result, as well as the calculated standard deviation.

Aiming to identify the influence of brucite decomposition and other phase transformations on the castables behavior, *in-situ* Young's modulus measurements were carried out during the first heating-cooling cycle of prismatic specimens (150 mm x 25 mm x 25 mm). These tests were conducted according to ASTM C 1198-91, using the resonance bar technique (Scanelastic equipment, ATCP, Brazil) in the temperature range of 30-1400°C and with a heating rate of 2°C/min. Further details of this technique can be found elsewhere [40].

As the incorporation of the drying agents into the compositions tends to increase the samples' permeability and pore volume content contained in the microstructure [33,35], assisted sinterability and corrosion measurements (via cup-tests) were also evaluated in this work to investigate the influence of such effects on the overall behavior of the $Al_2O_3$-MgO castables at high temperatures. Cylindrical samples (height and external diameter = 50 mm and central inner diameter = 12.5 mm) were prepared according to the 51053 DIN standard for the sinterability analyses. Such tests were carried out up to 1450°C for 5h, under a compressive load of 0.02 MPa and with a heating rate of 2°C/min in RUL testing equipment (Netzsch, Germany). The corrosion measurements were also conducted with cylindrical samples (with an external d = 50 mm, h = 50 mm and a central inner hole with 20 mm in diameter and 25 mm deep) pre-fired up to 1450°C for 5h. A total of 3 specimens were analyzed for each studied formulation. Before the tests, the inner cup of the castables were filled in with 8 g of a synthetic slag (Table 3) and the set (refractory + slag powder) was placed in an electrical furnace (Lindberg Blue, Lindberg Corporation, USA) and heated up to 1450°C for 2h with a heating rate of 2°C/min. After cooling, the corroded samples were cut and had their cross sections observed using the Image J 1.42q software (Wayne Rasband, National Institutes of Health, USA) for the determination of the slag penetration area. The latter parameter represents the relative infiltrated area of the castable. The original and the infiltrated areas of the sample cross section were measured and, after that, the percentage of the slag penetration was calculated, as described elsewhere [41].

Table 3 – Chemical composition of the synthetic slag used in the corrosion tests.

| Oxides (wt.%) | $SiO_2$ | $TiO_2$ | $Al_2O_3$ | $Fe_2O_3$ | CaO | MgO | $K_2O$ | $SO_3$ | F |
|---|---|---|---|---|---|---|---|---|---|
| Slag | 28.8 | 0.6 | 10.7 | 1.2 | 42.7 | 7.2 | 0.6 | 1.6 | 6.2 |

## 3. Results and discussion

*3.1 – Physical-chemical transformations derived from the interaction of the selected additives with MgO*

Firstly, mixtures of MgO + additives were prepared and analyzed in order to identify the most likely reactions that could take place when keeping the prepared samples at 30°C and/or 110°C for 24 hours. Although 30 wt.% of distilled water was added to MgO for the preparation of the MgO-$H_2O$ mixture (Table 1), this composition presented an aqueous layer (excess of liquid that did not react with magnesia) at the top of the molded samples after their curing step at 30°C for 24h. Hence, this unreacted liquid was discarded and only the solid and hydrated material was ground and analyzed using the TG and DSC techniques. This is the reason why a lower mass loss was observed for MgO-$H_2O$ mixtures, as presented in Fig. 1a and 1b and Table 4.

In general, all cured samples presented fast and high mass loss in the 30-150°C range due to the free-water release, and a second event (hydrate and additive decomposition) could also be detected in the 300-400°C one (Fig. 1a). Based on the DSC analyses, the most intense transformation for the cured compositions was the endothermic reaction associated with free-water release and two additional changes (small endothermic peak around 320-350°C, followed by an exothermic one) could also be identified in the samples' heating profiles (Fig. 1c). The latter reactions indicate that a limited amount of hydrated phases might be present in the materials kept at 30°C, as the decomposition of this phase absorbs energy, resulting in the endothermic transformation observed at 320-350°C. Besides that, the exothermic peak between 350-400°C may be associated with the desorption of the carboxylic acid molecules from the MgO particle's surface (Fig. 1c) and the decomposition of OAS (~375°C), as the gaseous products (CO and H$_2$O) derived from these transformations can later react with each other and give rise to an exothermic reaction called water-gas shift (Eq. 4) [13,42].

$$H_2O_{(g)} + CO_{(g)} \leftrightarrow CO_{2(g)} + H_{2(g)} \qquad \Delta H = -40.6 kJ/mol \qquad (4)$$

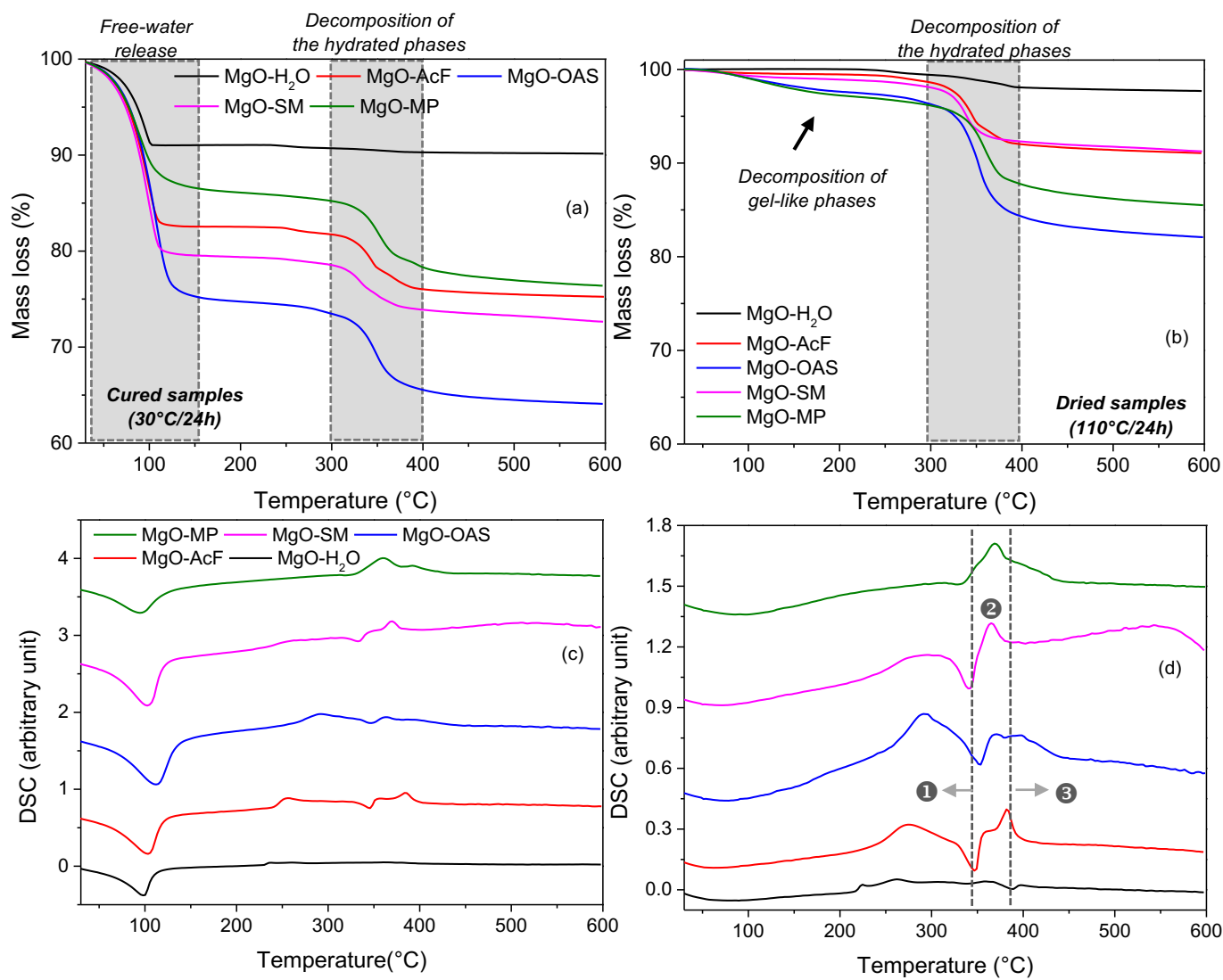

Fig. 1 – Mass loss and DSC profiles of the evaluated MgO-based compositions during their first heating treatment up to 600°C. The samples were kept at (*a* and *c*) 30°C/24h or (*b* and *d*) 110°C/24h before the measurements. The numbers pointed out in (*d*) represent: (1) hydrotalcite-like phases, Mg$_3$Si$_2$O$_5$(OH)$_4$ and/or Mg$_3$O(OH)$_4$ decomposition, (2) desorption of the carboxylic acid and OAS decomposition, and (3) brucite decomposition.

Table 4 – Total mass loss of the simplified compositions (MgO + additives) after the TG/DSC tests.

| Samples obtained after: | Total mass loss (%) after the TG/DSC measurements | | | | |
|---|---|---|---|---|---|
| | MgO-H$_2$O | MgO-AcF | MgO-OAS | MgO-SM | MgO-MP |
| Curing at 30°C/24h | 9.85 | 24.77 | 35.92 | 27.36 | 23.61 |
| Drying at 110°C/24h | 2.31 | 8.94 | 17.92 | 8.75 | 14.49 |
| ΔM (M$_{curing}$-M$_{drying}$) | 7.54 | 15.83 | 18.00 | 18.61 | 9.12 |

On the other hand, MgO-H$_2$O, MgO-AcF and MgO-SM compositions dried at 110°C for 24h and then subjected to the TG measurements (Fig. 1b) did not show the presence of free-water in their structure, whereas the mixtures comprised by MgO plus the organic aluminum salt (OAS) or the permeability enhancing active compound (MP) still presented a small mass change around 80-200°C, which might be related to the decomposition of gel-like phases [33,35] or the release of inter-lamellar water contained in hydrotalcite-like compounds [12], both of them derived from the action of these additives (as will be pointed out when discussing the XRD results). In fact, organic additives, such as OAS, may undergo hydrolysis in the first contact with water [43] and also favor the complexation of magnesium ions [20]. Such transformations are expected to prevent brucite formation and help to control the negative effects associated with magnesia hydration.

Fig. 2 shows the $^{27}$Al-NMR spectra obtained for MgO-OAS composition after curing, drying and firing steps. As the identified aluminum is derived from the Al-based salt incorporated into the evaluated mixture, one can infer that this compound was partially hydrolyzed, mainly generating the octahedral aluminum complex, $[(R-COO)Al(H_2O)_4]^{2+}$, in the cured and thermally treated samples up to 300°C (Fig. 2a). Moreover, four-coordinated aluminum units were also detected in the MgO-OAS composition above 200°C. An additional complex $[(R-COO)_2Al(H_2O)_2]^{+}$ and more intense peaks associated with four-coordinated aluminum units could be observed when increasing the firing temperature of the samples from 350°C up to 600°C (Fig. 2b). The intensity decrease of the $[(R-COO)Al(H_2O)_4]^{2+}$ peak and further identification of $[(R-COO)_2Al(H_2O)_2]^{+}$ ones indicates that decomposition of hydrated complex should take place in this temperature range, as also observed in the TG/DSC measurements.

Besides that, some attempts to perform $^{25}$Mg NMR analyses to evaluate the prepared compositions were carried out in this work, but no reliable results were obtained due to equipment limitations. Another aspect that makes it more difficult to identify the role of OAS in the presence of MgO and water is the fact that $Mg^{2+}$ may also generate hydrated compounds with octahedral coordination (similar to $Al^{3+}$), which inhibits a clear separation of the contribution of each of these ions. Hence, the combination of NMR spectroscopy and other techniques are required to better explain the phase transformations in the evaluated system.

Consequently, XRD measurements of all the dried samples were carried out (Fig. 3) and brucite was only detected as a minor phase in the $MgO-H_2O$ and MgO-OAS compositions. The addition of the organic aluminum salt (OAS) to the prepared MgO-OAS suspension also induced the generation of $Mg_3O(OH)_4$ (which is comprised by MgO and $Mg(OH)_2$ layers and may be retained in the mixture during the periclase conversion to brucite [30]) and $Mg_6Al_2(OH)_{18}.4.5H_2O$ / $Mg_6Al_2(OH)_{16}(CO_3).4H_2O$ (both compounds have similar main diffraction peaks) phases. The latter hydrates can be classified as hydrotalcite-like phases, as it shows a $Mg_xAl_y(OH)_{2x+3y}.nH_2O/Mg_xAl_y(OH)_{2x+2y}(CO_3)_{y/2}.nH_2O$ stoichiometry. As reported in the literature, hydrotalcites can be obtained by $Al(OH)_3$ and $Mg(OH)_2$ reaction in aqueous medium [29,44] and, as the MgO-OAS suspensions contained $[(R-COO)Al(H_2O)_4]^{2+}$ derived from the organic salt partial hydrolysis, as well as MgO in its non-hydrated and hydrated form, further interaction of these species led to $Mg_6Al_2(OH)_{18}.4.5H_2O$ / $Mg_6Al_2(OH)_{16}(CO_3).4H_2O$ generation in the evaluated processing conditions.

Hydrotalcite-like phases were also the only crystalline hydrates found in the MgO-MP composition (Fig. 3e), although a magnesium hydroxi-silicate $[Mg_3Si_2O_5(OH)_4]$ could be formed in the MgO-SM sample, due to the presence of silica fume in SioxX-Mag and its further interaction with magnesia and water. According to Table 4, higher ΔM (total mass loss difference between cured and dried samples) values for MgO-AcF, MgO-LA and MgO-SM indicate that a greater content of hydrated compounds (crystalline or gel-like ones) could be formed in these suspensions after keeping them at 110°C.

Regarding the DSC results of the dried MgO-based suspensions (Fig. 1d), more intense endothermic and exothermic peaks could be detected in the 300-390°C range, as a consequence of the higher content of hydrated phases derived from the further interaction of MgO with water vapor at 110°C. Although additional tests must be carried out to establish the decomposition sequence of the hydrated compounds present in these samples, it is believed that the small endothermic peak identified in the profiles of $MgO-H_2O$ (reference) and MgO-OAS at ~380°C (Fig. 1d) should be associated with the brucite phase (as only these two systems presented $Mg(OH)_2$ in its composition, Fig. 3). On the other hand, the first endothermic transformation in the 300-360°C range might be related to $Mg_3O(OH)_4$, $Mg_6Al_2(OH)_{18}.4.5H_2O$ / $Mg_6Al_2(OH)_{16}(CO_3).4H_2O$ and/or Mg$_3$Si$_2$O$_5$(OH)$_4$ decomposition, and an exothermic peak points out the desorption of the carboxylic acid molecules from the MgO particle's surface and OAS decomposition around 360-390°C (Fig. 1d).

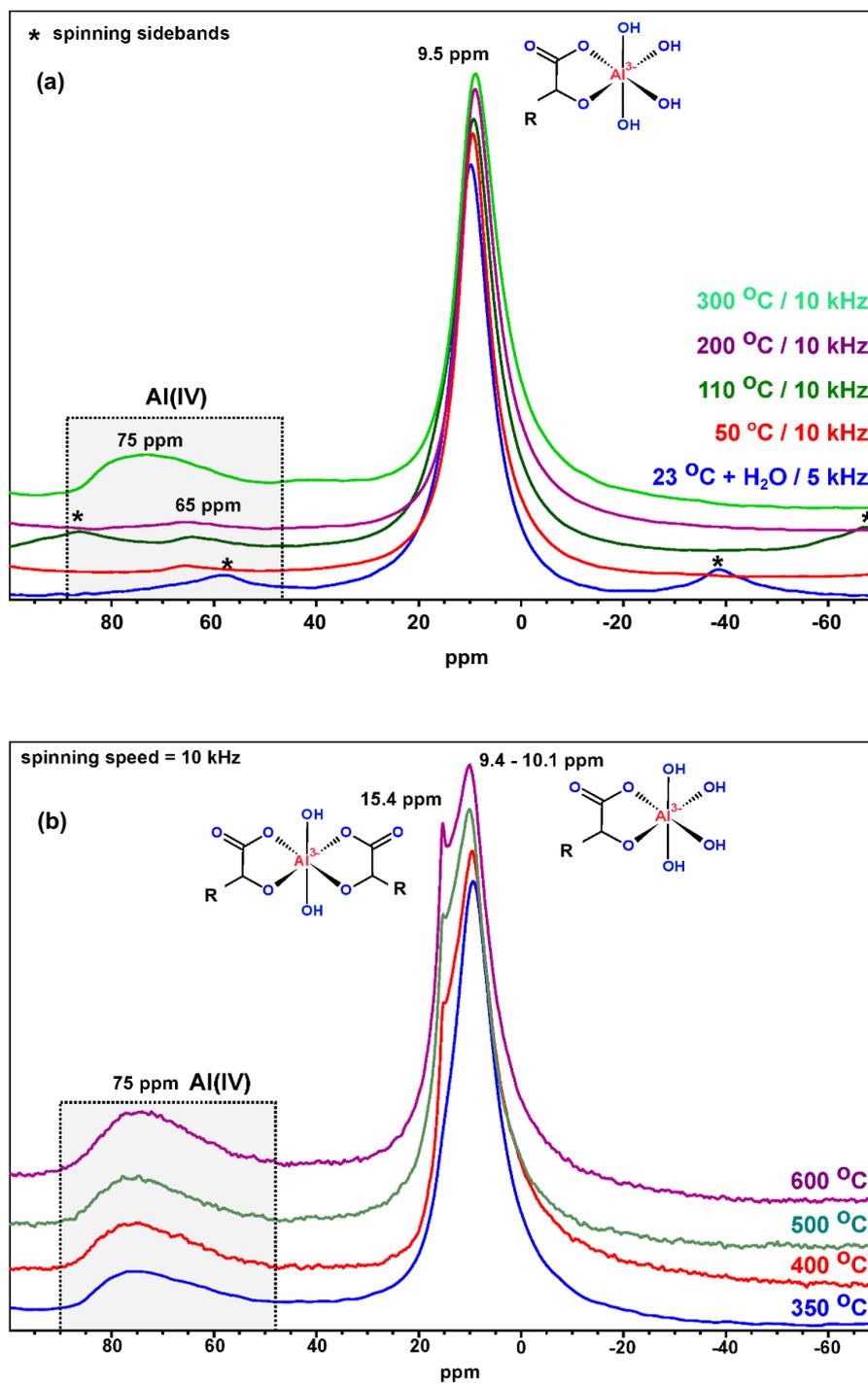

Fig. 2 – $^{27}$Al-ssNMR spectra of MgO-OAS composition after different thermal treatments: (a) curing (30°C or 50°C for 24h), drying (110°C for 24h) or firing at 200°C or 300°C for 2h, and (b) firing in the range of 350°C and 600°C for 2h.

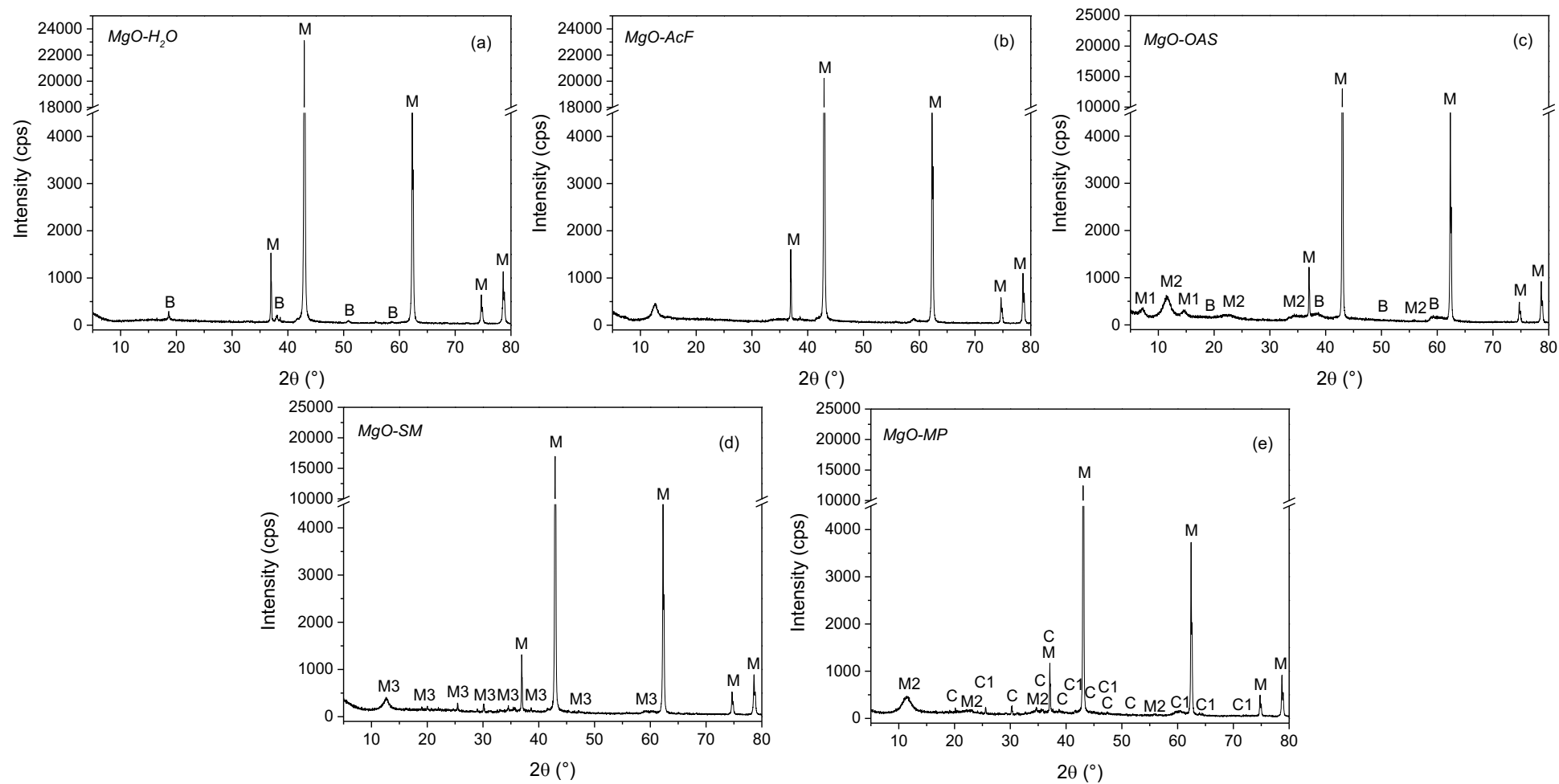

Fig. 3 – X-ray diffraction profiles of MgO-based compositions kept at 110°C for 24h. The additives incorporated into the magnesia-based suspensions were: formic acid - AcF (b), organic aluminum salt - OAS (c), SioxX-Mag – SM (d) and Refpac MIPORE 20 – MP (d). The results obtained for MgO-H$_2$O (reference) is shown in (a). M = MgO, B = Mg(OH)$_2$, M1 = Mg$_3$O(OH)$_4$, M2 = Mg$_6$Al$_2$(OH)$_{18}$·4.5H$_2$O / Mg$_6$Al$_2$(OH)$_{16}$(CO$_3$)·4H$_2$O (both phases have similar XRD patterns), M3 = Mg$_3$Si$_2$O$_5$(OH)$_4$, C = CaAl$_2$O$_4$ and C1 = CaAl$_4$O$_7$.

As observed in the results presented in Fig. 1 and 3, it is not a straightforward task to clearly separate the decomposition and XRD peaks associated with each hydrated phase derived from MgO hydration and its interaction with OAS, due to the complexity of this system. Consequently, different phase transformations are expected to take place in the resulting microstructure of high-alumina MgO-bonded castables, when adding the selected additives presented above to such formulations. The following sections highlight the influence of such materials in various properties of the designed vibratable castables.

*3.2 – Water demand, flowability and curing behavior of the evaluated castables*

Fig. 4 shows the water content and vibra-flow values obtained for the $Al_2O_3$-MgO castables evaluated in this work. 3.9 up to 4.2 wt.% of water was required to prepare the designed formulations (except RefM-MP) in order to obtain flowability values higher than 150%. The incorporation of SioxX®-Mag into RefM-SM enhanced the vibra-flow level (~182%) of this castable, which points out the good influence of this alumina-silica based additive in such a property [37]. The opposite trend was observed for the RefM-MP composition, although this other drying agent required higher liquid content (4.7 wt.%) to result in a homogeneous mixture with vibra-flow of 156%. Considering that Refpac MIPORE® 20 contains $CaAl_2O_4$ and $CaAl_4O_7$ in its composition (as detected in Fig. 3) and other organic components (including dispersants) [35], it was expected that some of these components might also react with water and partially consume the available liquid. Consequently, RefM-MP presented a water demand increase above 20% when compared to the reference castable (RefM). For RefM-PF and RefM-OAS compositions, just small changes in the liquid content (4.1-4.2 wt.%) were observed and both compositions reached the desired vibra-flow level.

The addition of drying agents to the prepared castables led to significant differences in their curing behavior. As highlighted in Fig. 5, a fast increase in the ultrasonic wave propagation velocity for the fresh reference refractory (drying additive-free) could be detected up to 4 h after the beginning of the test and the complete stiffening of the mixture most likely took place after ~10 h (maximum velocity). Additionally, a further drop in the measured ultrasonic velocity could also be detected for RefM when considering longer curing times (> 11h), which indicated that cracks and flaws might be formed in the resulting microstructure as a consequence of the brucite phase generation.

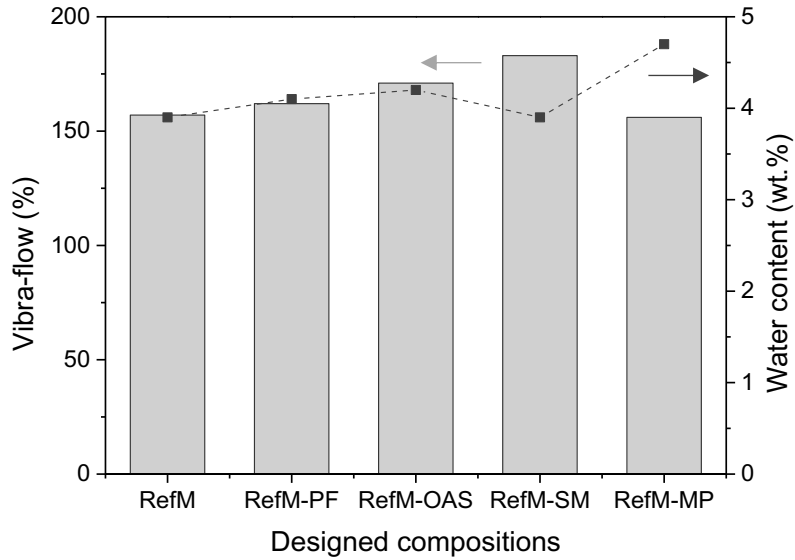

Fig. 4 – Vibratable flow and water demand for the preparation of the designed high-alumina MgO-bonded castables containing different drying agents.

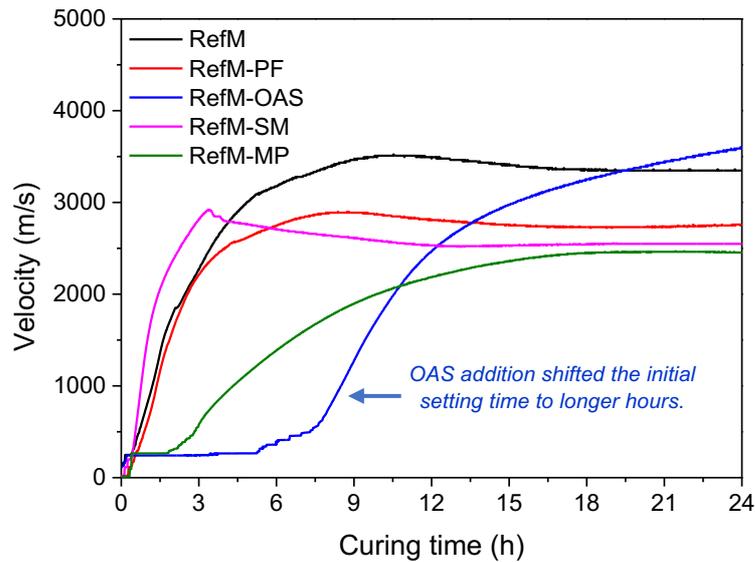

Fig. 5 – Curing behavior of the fresh castable mixes via ultrasonic wave propagation velocity evaluations at 22°C and for 24h.

As expected, RefM-PF showed a similar velocity profile during curing as the one obtained for RefM (Fig. 5), but the presence of the polymeric fibers and the slightly higher water content added to the former resulted in lower velocity results after reaching 24 h (2759 m/s instead of 3347 m/s). On the other hand, the other evaluated drying agents induced major changes in the setting time of the castables. For instance, SM sped up the stiffening of the REFM-SM composition due to the interaction of silica, magnesia and water contained in this system, which should have favored the formation of the chrysotile-like compound [$Mg_3Si_2O_5(OH)_4$, as identified in the X-ray diffraction analyses, Fig. 2] and other magnesium-silicate-hydrated gels [24,27]. The highest velocity value for RefM-SM composition was observed around 3h and an additional drop of this property (most likely due to the sample's cracking) could be identified between 3 to 16 h of curing (Fig. 5).

Both compounds MP and OAS led to a delay of the castables' setting time, where the latter shifted the beginning of the ultrasonic velocity increase to ~7h. After that, a continuous rising of this property still took place and its maximum value was not even observed up to 24h. As discussed in Section 3.1, MP or OAS reaction with MgO and water results in the formation of $Mg_3O(OH)_4$, $Mg_6Al_2(OH)_{18} \cdot 4.5H_2O$/ $Mg_6Al_2(OH)_{16}(CO_3) \cdot 4H_2O$ and gel-like phases generation at low temperatures [33,35], inhibiting or decreasing the amount of brucite to be formed in the resulting microstructure. Therefore, these changes in the MgO hydration sequence [giving rise to additional hydrated compounds and not only $Mg(OH)_2$] might be pointed out as being responsible for the longer setting time of RefM-MP and RefM-OAS compositions.

*3.3 – Cold flexural strength and apparent porosity*

Keeping in mind that the green mechanical strength of the refractories may play a role in their explosion likelihood, as spalling should mainly take place when; (*i*) the resulting steam pressure built up overcomes the bond strength derived from the binder action [1,30,47] or (*ii*) the temperature gradient between the surface and inner region of the ceramic lining leads to thermomechanical stresses, which combined with the steam pressure, induce its failure; the cold flexural strength and apparent porosity of the prepared MgO-bonded castables were analyzed after the curing (30°C for 24h) and drying (110°C for 24h) steps. Fig. 6a shows that RefM-OAS presented the best performance in both tested conditions and a significant enhancement of the samples' modulus of rupture (reaching ~5.7 ± 0.8 MPa) could be observed after drying. This behavior can be associated with the progress of the hydration reactions that may still occur at 110°C due to the exposure of the MgO contained in the castable microstructure to water vapor [12,30]. In this case, a suitable accommodation of the brucite crystals in the available pores of the RefM-OAS castable led to the enhancement of its mechanical properties.

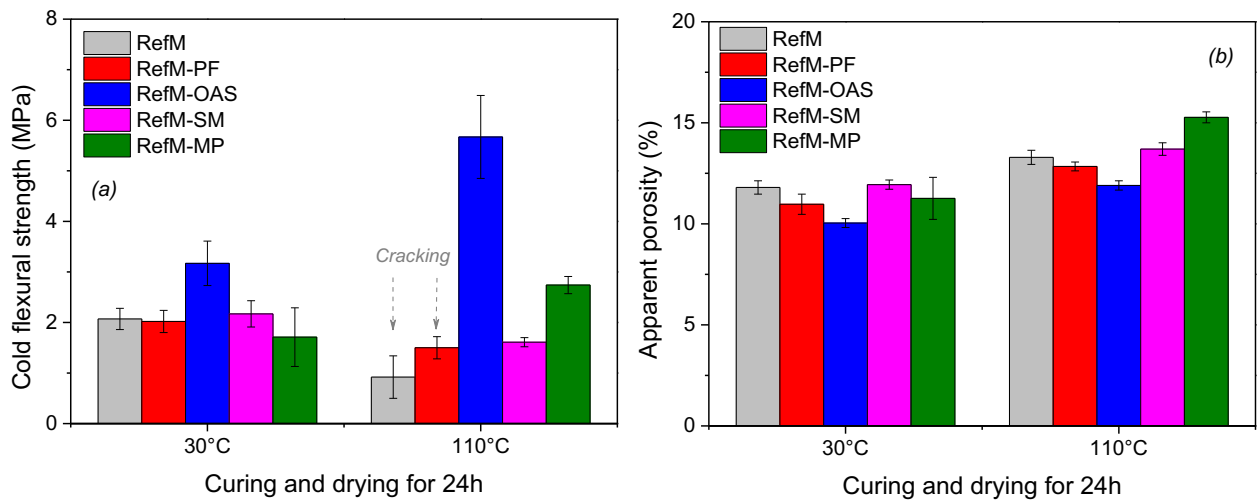

Fig. 6 – (a) Cold flexural strength (3-point bending test) and (b) apparent porosity of the evaluated castables after curing (30°C for 24h) and drying at 110°C for another 24h.

The same process (hydrates generation during drying) in RefM and RefM-PF compositions had a detrimental effect in their mechanical strength (Fig. 6a), as the non-accommodation of the novel brucite crystals in the resulting structure led to the formation of small superficial cracks on the obtained samples. Formic acid was incorporated into the castables preparation, as this compound may adsorb on MgO particles and decrease the energetic barrier for the interaction of this oxide with water, resulting in a greater likelihood for nuclei formation and, consequently, leading to a more controlled growth of the brucite crystals [12]. Despite the influence of this organic compound in the behavior of the RefM refractory, its action was not efficient enough to prevent cracking at 110°C. Compared to the refractories produced by Santos Jr. et al. [12], the only difference in the processing conditions used in the present work was the new batch of the dead-burnt magnesia, also provided by RHI-Magnesita. Due to the higher MgO reactivity of this new batch, most likely a further adjustment in the formic acid (increase in its content) would be required to obtain crack-free samples.

Although no flaws were identified on the surface of the RefM-SM castable, this composition also presented a drop in its modulus of rupture after drying at 110°C, which indicates that the expansive feature of magnesia hydration reaction negatively affected this property. All castables presented a small apparent porosity increase when going from curing to drying conditions (Fig. 6b), due to free-water release and the likely partial decomposition of gel-like phases of the samples' structure.

*3.4 – Drying and explosion tests*

Cured cylindrical samples (d = 50 mm x h = 50 mm, kept at 30°C for 24h) were subjected to thermogravimetric measurements with different heating rates to analyze their drying behavior and explosion resistance. Fig. 7 shows the mass loss and DTG profiles of castables in the 30-600°C range, under heating rates of 2, 5 or 20°C/min. When applying the most severe heating condition (20°C/min, Fig. 7a and 7b), three refractory systems exploded (RefM, RefM-PF and RefM-SM) and only the samples containing OAS or MP survived with no spalling. The samples' explosion was identified around 124-157°C in Fig. 7a, which are relatively low temperatures, considering that brucite decomposition only takes place at ~380-420°C. Thus, two main factors might be responsible for the earlier explosion of the RefM, RefM-PF and RefM-SM castables: low permeability and/or reduced green mechanical strength (as observed in Fig. 6a).

As the evaluated samples were only cured before testing, a large amount of water was still present in the microstructure during their first heating step (Fig. 7a). Such an aspect combined with the fast heating rate (20°C/min) and most likely low permeability of the castables shifted the water evaporation and ebullition to higher temperatures, resulting in higher steam pore pressure and greater spalling risk. As a consequence, the green mechanical strength of RefM, RefM-PF and RefM-SM compositions was not enough to withstand the thermomechanical stress derived from these transformations (Fig. 7a). The polymeric fibers did not prevent RefM-PF explosion during fast heating (20°C/min) because, despite their low melting point (~104°C), such material should mainly act enhancing the refractory permeability after its full decomposition (> 300°C) [35,39]. On the other hand, the role of SioxX®-Mag in optimizing the properties of MgO-containing compositions is based on the precipitation of crystalline and gel-like hydrated phases that help control brucite formation [26,45]. These amorphous hydrated compounds usually fill in the pores of the castable's structure, which may significantly reduce its permeability and, consequently, the steam withdrawal during heating.

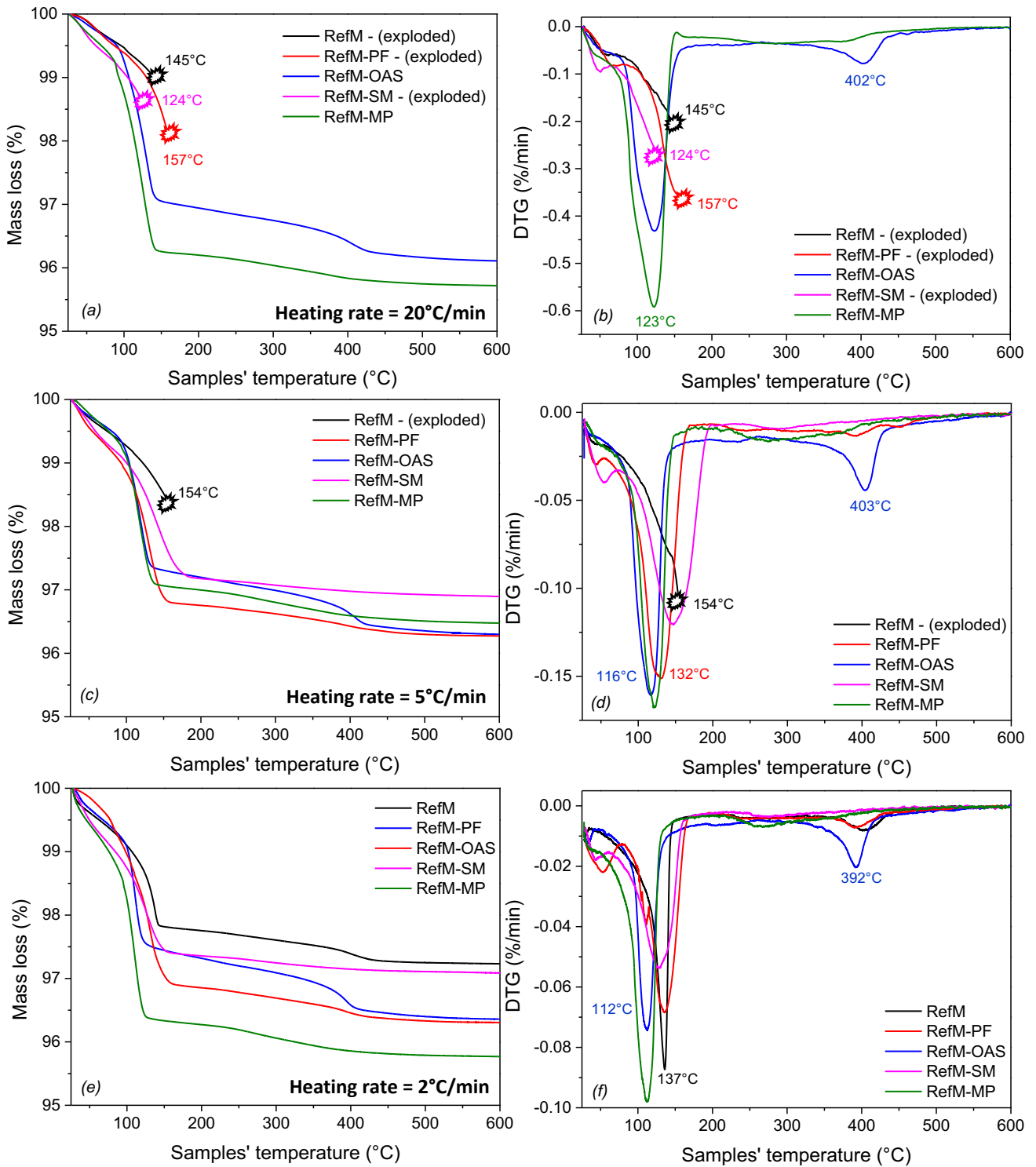

Fig. 7 – (a, c and e) Mass loss (%) and (b, d and f) first derivative of the mass loss curve for the thermal treatments carried out with a heating rate of 20, 5 and 2 °C/min. All samples were cured at 30°C for 1 day before the measurements.

The DTG profiles shown in Fig. 7b indicate that both RefM-MP and RefM-OAS were able to expel a higher amount of water in the 80-123°C range (more intense peak), indicating that most likely the hydrotalcite-like phases derived from the action of these two additives allowed inter-lamellar water withdrawal at low temperatures or likely the gel-like phases underwent decomposition at such conditions, giving rise to a great number of permeable paths in the resulting microstructure [35,46]. Additionally, another interesting aspect is that the RefM-OAS sample also presented a further broad decomposition range (peak around 402°C), which can be related to $Mg(OH)_2$, hydrotalcite-like phases and OAS decomposition (Fig. 7b).

This decomposition peak close to 400°C could still be detected for RefM-OAS refractory when the TG tests were carried out with heating rates of 5 or 2°C/min (Fig. 7d e 7f). Besides that, all the designed compositions withstood the thermomechanical stresses under a more conservative heating procedure (2°C/min, Fig. 7e), but the resultant steam pressure in the RefM microstructure led to its explosion at 154°C when the measurements were carried out at 5°C/min. Therefore, the selected drying agents helped to improve the performance of the designed $Al_2O_3$-MgO castables mainly when applying heating rates of 5 or 20°C/min. The most promising additives were MP and OAS, as RefM-MP and RefM-OAS cured samples were able to overcome severe drying conditions without any spalling effect.

Despite the apparent good performance of the reference composition during the slower heating procedure (2°C/min), the collected samples showed some surface cracks after the drying experiments, as highlighted in Fig. 8. Thus, as expected, these results confirm that MgO-bonded compositions have a higher trend to develop cracks or even explode during their first thermal treatment, which requires using additional compounds to adjust their microstructure and inhibit such undesirable flaws.

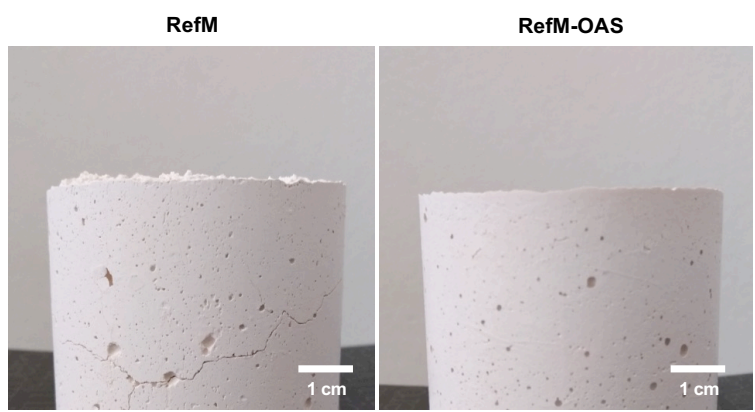

Fig. 8 – Images of the castables' samples obtained after the thermogravimetric tests carried out with a heating rate of 2°C/min and up to 600°C.

*3.5 – Characterization of the castable samples at high temperatures*

In order to identify the temperature influence on the stiffness evolution of the $Al_2O_3$-MgO castables, hot Young's modulus measurements were carried out in dried samples (110°C for 24h) in the 30-1400°C temperature range and during heating-cooling steps. In tune with the previous flexural strength analyses (Fig. 6a), RefM and RefM-PF presented the lowest E values at the beginning of the experiments as a consequence of the small cracks located on their surfaces (Fig. 9). A continuous drop of the samples' stiffness was identified when heating them from 30 to 400°C, which is mainly related to hydrates and additive decomposition. After that, the following main transformations that took place in the resulting microstructure are: *(i)* coarsening and densification of the small MgO crystals (> 800°C) derived from the hydrated compound decomposition [30], and *(ii)* sintering mechanisms and $MgAl_2O_4$ generation above 1000°C due to MgO and $Al_2O_3$ reaction [1,27,30,48]. No further E changes were observed for RefM and RefM-PF during cooling from 1400°C down to room temperature, but an increment of ~150% of the sample's stiffness was obtained after a full thermal cycle, when comparing the initial and final Young's modulus values.

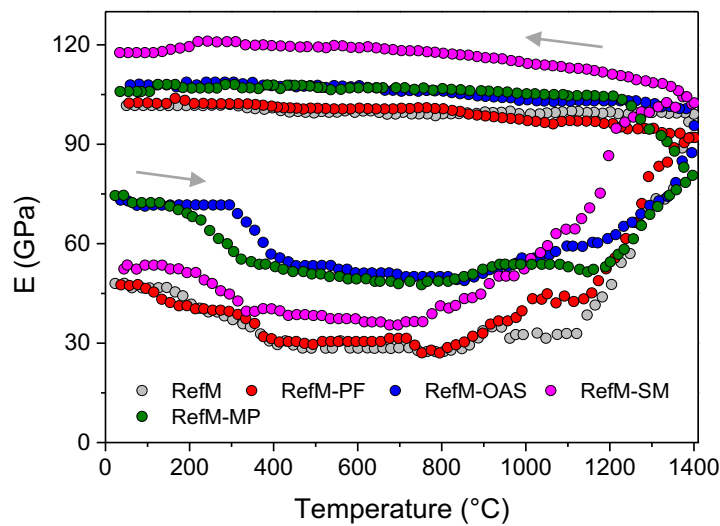

Fig. 9 – Young's modulus evolution as a function of the temperature of the dried samples of the designed castables during their first heating-cooling cycle (30-1400°C). The measurements were carried out with a heating rate of 2°C/min.

RefM-OAS and RefM-MP presented similar elastic modulus results during one full thermal cycle, where the main difference of them and the reference refractory is the fact that they already have enhanced stiffness at the beginning of the E measurements (~75 GPa, Fig. 9). The hydrate decomposition could also be inferred by the hot Young's modulus analyses during heating of the OAS and MP-containing castables, which resulted in the decay of this property in the 200-400°C range.

RefM-SM, on the other hand, was the composition that presented a faster and earlier rise of its E level during heating (Fig. 9). Such a performance may be induced by a small amount of liquid phase formation in the refractory microstructure (due to silica fume interaction with MgO, $Al_2O_3$ and their impurities) at high temperatures, providing a more effective ions diffusion and, consequently, enhancing the microstructure densification. A positive aspect of these phase transformations in RefM-SM samples is the likelihood of producing a refractory with improved Young's modulus (high E values after firing), as shown in Fig. 9.

However, it is important to highlight that *in situ* spinel formation should also take place in such composition and this transformation is commonly followed by an expansion attributed to the density differences among the reactants and the product (MgO - 3.58 $g/cm^3$, $Al_2O_3$ - 3.98 $g/cm^3$ and $MgAl_2O_4$ - 3.60 $g/cm^3$), leading to a volumetric expansion close to 8% and a linear one of 2.6% for an equimolar alumina-magnesia mixture [1,48,49]. Fig. 10 indicates that RefM-SM presented an overall linear expansion of approximately 2.7% after keeping this sample at 1450°C for 5h, whereas the other castables reached 1.5-1.8% at the same conditions. Previous works [48,50] reported that high $MgO/SiO_2$ wt.% ratio (> 12) in castables' matrix may be detrimental, resulting in high permanent linear changes (> 2%) and cracking of the material. On the other hand, low ratio values (< 3) are also not suitable due to significant shrinkage and flaw generation. Hence, the most common $MgO/SiO_2$ values used in industrial applications are between 4 and 8. RefM-SM contained 6 wt.% of MgO and 0.35 wt.% of $SiO_2$, resulting in a magnesia-silica ratio of 17, which explains the high $dL/L_0$ results shown in Fig. 10. Nevertheless, despite the high expansion of this refractory, it seems that spinel phase was properly accommodated in the resulting microstructure, as no visual cracks or flaws could be identified on the surface of the fired samples. Additionally, RefM-SM presented the highest Young's modulus values among the tested materials after one heating-cooling cycle (Fig. 9).

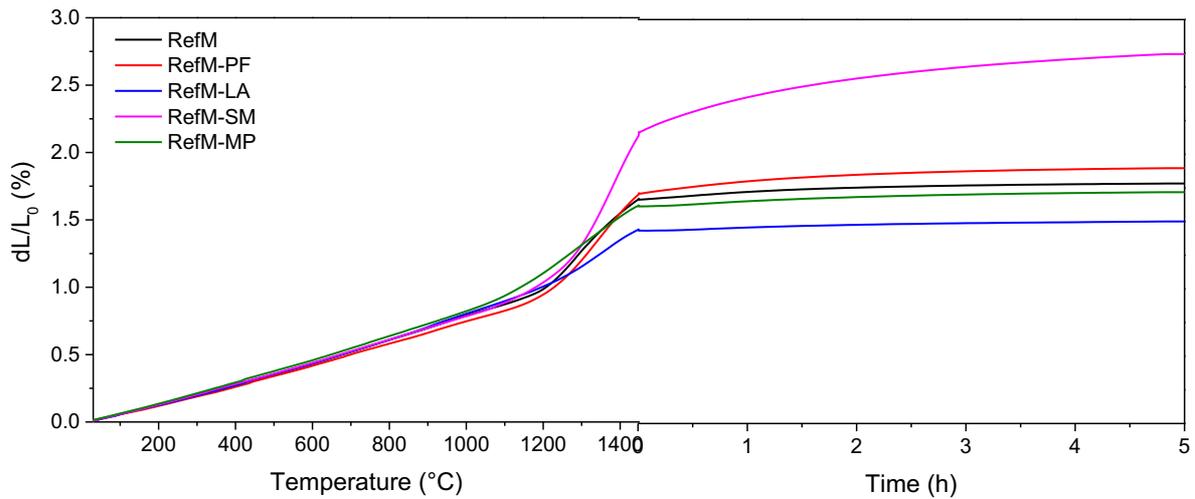

Fig. 10 – Expansion behavior of the designed $Al_2O_3$-MgO castables containing different drying agents. The measurements were carried out with a heating rate of 2°C/min, up to 1450°C for 5h and using an applied load of 0.02 MPa.

As the evaluated drying additives have the ability to induce transformations that may change the porosity and permeability level of the refractories' microstructure, an additional experimental test was carried out in this work (corrosion cup-test), aiming to analyze whether the action of such compounds could affect the slag infiltration and corrosion resistance of the prepared castables. Table 5 presents the calculated slag infiltrated area for each composition.

Table 5 – Slag penetration area calculated for each designed castable after corrosion tests at 1450°C for 2h.

| Castables | RefM | RefM-PF | RefM-OAS | RefM-SM | RefM-MP |
|---|---|---|---|---|---|
| Slag penetration (%) | 16.45 ± 0.94 | 30.28 ± 0.97 | 12.61 ± 1.95 | 21.97 ± 0.74 | 16.98 ± 1.39 |

The interaction of the molten slag with the $Al_2O_3$-MgO castables at 1450°C led to slag infiltration in their microstructures, where the reference material had ~16.45% of its area penetrated by the liquid. The addition of PF or SM to the designed reference composition affected the overall corrosion resistance and dimensional stability of the obtained refractories, as not only RefM-PF showed higher slag infiltration (~30.28%) with the liquid phase reaching the side external surface of the samples, but also RefM-SM (21.97%) followed by expansion and large cracks on the external surface of the samples' upper region. Hence, the presence of $SiO_2$ in the latter refractory seems to be detrimental to its corrosion behavior. On the other hand, MP and OAS seem to be the most suitable drying agents to $Al_2O_3$-MgO castables, as the phase transformations induced by them did not change or even resulted in lower liquid infiltration in the specimens' microstructure (Table 5).

## 4. Conclusions

Designing and processing crack-free high-alumina MgO-bonded castables are huge challenges for refractory producers, as not only cracks and flaws can be generated during the curing step of the prepared samples (due to brucite formation and its non-accommodation in the microstructure), but also the first heating treatment of such materials may lead to their spalling or explosion, as a consequence of a combination of low permeability and high steam pore pressure build-up in the ceramic lining. According to the results presented in this work, selecting and using suitable drying agents may be an important alternative to optimize the properties and develop refractories with improved performance and high explosion resistance. The incorporation of an organic aluminum salt (OAS) or a permeability enhancing active compound (MP) to the design vibratable castable helped to inhibit the samples' explosion even under severe heating conditions (2, 5 or 20°C/min), increasing their green mechanical strength and slag infiltration resistance when compared to the additive-free composition. Thus, these drying agents (OAS and MP) are the most recommended ones to be added to $Al_2O_3$-MgO refractories.

On the other hand, polymeric fibers (PF) and a silica-based compound (SM) were not able to prevent the explosion of the RefM-PF and RefM-SM castables when using a heating rate of 20°C/min and other parallel negative aspects (samples' cracking during drying at 110°C, high dimensional linear expansion and increased slag penetration during corrosion tests) could also be observed when testing these materials.

A forthcoming paper by the authors should discuss in more details (based on experimental and simulation data) the microstructural changes derived from OAS and MP role in $Al_2O_3$-MgO compositions during the first heating treatment of the castables, to better explain how they may influence the permeability and drying performance of such refractories.

## 5. Acknowledgements

The authors would like to thank Fundação de Amparo a Pesquisa do Estado de São Paulo (grant # 2019/07996-0) for supporting this work. The authors are also grateful to Almatis, RHI-Magnesita, Imerys Aluminates S.A. and Elkem for supplying the raw materials used in this study.